\documentstyle[prl,aps,psfig]{revtex}

\begin{document}
%\twocolumn[\hsize\textwidth\columnwidth\hsize\csname@twocolumnfalse\endcsname 

\title{Supernova Observation Via Neutrino-Nucleus Elastic Scattering 
in the CLEAN Detector}

\author{C.~J.~Horowitz
\footnote{e-mail:  horowitz@iucf.indiana.edu} 
}
\address{Nuclear Theory Center and Dept. of Physics, Indiana
University, Bloomington, IN 47405}

\author{K.~J.~Coakley
\footnote{Contributions of NIST staff to this work are not subject to copyright laws in the US.}
}
\address{National Institute of Standards and Technology, Boulder, CO 
80305}

\author{D.~N.~McKinsey}
\address{Physics Department, Princeton University, Princeton, NJ 08544}

\date{\today} 
\maketitle 
\begin{abstract}
Development of large mass detectors for low-energy neutrinos and dark matter may allow supernova detection via neutrino-nucleus elastic scattering.  An elastic-scattering detector could observe a few, or more, events per ton for a galactic supernova at 10 kpc ($3.1 \times 10^{20}$ m).  This large yield, a factor of at least 20 greater than that for existing light-water detectors, arises because of the very large coherent cross section and the sensitivity to all flavors of neutrinos and antineutrinos.   An elastic scattering detector can provide important information on the flux and spectrum of $\nu_\mu$ and $\nu_\tau$ from supernovae.  We consider many detectors and a range of target materials from $^4$He to $^{208}$Pb.  Monte Carlo simulations of low-energy backgrounds are presented for the liquid-neon-based Cryogenic Low Energy Astrophysics with Noble gases (CLEAN) detector.  The simulated background is much smaller than the expected signal from a galactic supernova.     
\end{abstract}
\vskip2.0pc
%]
\section{Introduction}

Rich information on neutrino properties, oscillations, the supernova 
mechanism, and very dense matter is contained in the neutrinos from 
core-collapse supernovae\cite{supernova}.  Existing detectors such as 
Super-Kamiokande\cite{SK} should accurately measure the $\bar\nu_e$ 
component of the supernova signal.  However the very interesting 
$\nu_\mu$, $\nu_\tau$, $\bar\nu_\mu$, and $\bar\nu_\tau$ 
(collectively $\nu_x$) components may be detected without direct 
energy information and or in the presence of significant backgrounds 
from other neutrino induced reactions.  Therefore, additional $\nu_x$ 
detectors could be very useful.

Perhaps the ``ultimate" supernova detector involves neutrino-nucleus 
elastic scattering \cite{elastic,cabrera}.  The count rate in such a 
detector could be very high because the coherent elastic cross 
section is large and all six neutrino components 
($\nu_e$,$\bar\nu_e$, and the four $\nu_x$) contribute to the 
signal. In particular, the detector is sensitive to $\nu_x$, which 
are expected to have a high energy and large cross section.  Elastic 
scattering detectors can have yields of a few or more $\nu_x$ events 
{\it per ton} for a supernova at 10 kpc ($3.1\times 10^{20}$ m).  
This is an increase by a factor of 20 or more over existing light-water detector yields of hundreds of $\bar\nu_e$ and tens of $\nu_x$ events {\it per kiloton}.  

Furthermore, the energy of nuclear recoils provides direct 
information on the $\nu_x$ spectrum.  Existing detectors measure 
$\nu_x$ via neutral-current inelastic reactions on oxygen\cite{vogel}, deuterium \cite{SNO}, or carbon \cite{Kamland}.  Here the observed energy deposition does not depend on the neutrino energy as long as it is above threshold.  Perhaps neutrino-proton elastic scattering\cite{beacom} can be detected in KamLAND\cite{Kamland}.  This is similar to neutrino-nucleus elastic scattering but has a smaller cross section.  

Alternatively, it may be possible to detect $\nu_x$ using inelastic 
excitations of Pb.  Proposals include using lead perchlorate, as 
suggested by Elliott\cite{elliott}, OMNIS \cite{omnis} and LAND 
\cite{land}.  Here some information on $\nu_x$ energies may be 
obtained by measuring the ratio of single- to two-neutron knockout.  
However, the inelastic Pb cross sections are somewhat uncertain.  In contrast, neutrino-nucleus elastic cross sections can be calculated accurately with very little theoretical uncertainty.

The $\nu_x$ spectrum depends on how neutrinos thermalize with matter in a supernova, and is somewhat uncertain.  Keil, Raffelt, and Janka have studied the effects of NN bremsstrahlung, pair annihilation, and nucleon recoil on the $\nu_x$ spectrum \cite{raffelt}.  These effects can be measured with an elastic-scattering detector.
 
Obtaining direct information on $\nu_x$ energies may be very 
important because the difference in energies for $\nu_x$ 
compared to $\nu_e$ or $\bar\nu_e$ is the primary lever arm for 
observing neutrino oscillations.  For example, 
$\nu_x\rightarrow\nu_e$ oscillations could lead to high-energy 
$\nu_e$.  However, deducing the oscillation probability may depend 
crucially on knowing how hot the $\nu_x$ were to begin with. 
Neutrino-nucleus elastic scattering itself is ``flavor blind".  
Therefore, the signal should be independent of neutrino oscillations 
(among active species).  Thus elastic scattering may provide a 
baseline with which to characterize the supernova source.  Comparing this 
information to other flavor-dependent information and theoretical 
simulations may provide the best evidence of oscillations.    

The kinetic energy of the recoiling nuclei is low, typically below 100 
keV.  It is difficult to detect such low-energy events in the 
presence of radioactive backgrounds.  Furthermore, scintillation 
signals from the nuclear recoils may be reduced by quenching because 
of the very high ionization density.  However, recent progress in 
designing detectors for low-energy solar neutrinos suggests that 
detection may be feasible.  In general, backgrounds for solar 
neutrinos with a low count-rate signal may be more severe than 
those for a supernova, where all of the events are concentrated in a 
few-second interval.  

Cryogenic Low Energy Astrophysics with Noble gases (CLEAN) is a proposed detector for low-energy solar neutrinos based on scintillation in an ultrapure cryogenic liquid\cite{Ne}.  This will 
detect electrons from neutrino-electron scattering at energies 
comparable to the recoil energy of nuclei from supernova neutrinos.  
In this paper, we discuss the utility of CLEAN for supernovae 
detection via neutrino-nucleus elastic scattering.      

There is considerable interest in the direct detection of weakly 
interacting massive particles (WIMPS).  These are expected to 
produce recoiling nuclei with a spectrum somewhat similar\cite{spectrum} to that of supernova neutrino-nucleus elastic scattering.  Again, backgrounds for WIMP detection may be larger than 
for supernovae because of the low WIMP count rates.  Present WIMP 
detectors, for example CDMS\cite{CDMS}, have small target masses.  
However, future detectors may be larger.

It is important to search for neutrinoless double-beta decay, as this 
can distinguish Dirac from Majorana neutrinos.  Existing $^{76}$Ge 
experiments use multikilogram masses \cite{hm}. The next-generation 
experiments such as Majorana \cite{majorana} or Genius \cite{genius} 
may employ up to a ton of Ge.  The need for good energy resolution, to 
tell neutrinoless from two-neutrino double-beta decay, often aids in 
the detection of low-energy recoils.  We calculate that the largest 
double-beta decay experiments may soon be sensitive to galactic supernovae 
via elastic scattering.

Finally, micropattern gas detectors \cite{mgd} may have a threshold low enough to detect nuclear recoils.  This may allow the observation of neutrino-nucleus elastic scattering using reactor antineutrinos.      
 
Thus the technical requirements for detecting low-energy solar neutrinos, WIMPS, 
double-beta decay, and supernovae via nuclear-elastic scattering may 
be similar.  One detector or approach for low-threshold, low-background, large mass measurements may have applications in multiple areas including supernova detection via neutrino-nucleus scattering.   

In section II, we fold elastic scattering cross sections with a model 
supernova neutrino spectrum to produce recoil spectra and yields.  We 
consider a range of noble-gas targets from $^4$He to $^{132}$Xe along 
with $^{12}$C, $^{28}$Si, $^{76}$Ge, $^{114}$Cd, $^{130}$Te, and 
$^{208}$Pb.  We also discuss extrapolating yields to nearby 
isotopes.  Section III focuses on the liquid-Ne-based CLEAN 
detector, which appears to be very promising.  A Monte Carlo 
simulation of backgrounds in CLEAN is presented and compared to the 
expected supernova signal.  We discuss the large signal-to-background 
ratio, choice of fiducial volume, and possible detector thresholds.  
We conclude in section IV.

\section{Supernova Signals in Various Detectors}

The neutrino-nucleus elastic-scattering cross section 
$d\sigma/d\Omega$ is \cite{freedman,elastic},
\begin{equation}
{d\sigma\over d\Omega} =   {G^2\over 4\pi^2} k^2 (1+{\rm cos}\theta){ 
Q_w^2\over 4} F(Q^2)^2,
\end{equation}
for a neutrino of energy $k$ scattering at angle $\theta$.  The Fermi constant is $G$.  This 
coherent cross section depends on the square of the weak charge $Q_w$
\begin{equation}
Q_w=N - (1 - 4{\rm sin}^2\Theta_W) Z
\label{qw} 
\end{equation}
of a nucleus with $N$ neutrons and $Z$ protons.  The weak mixing 
angle is sin$^2\Theta_W\approx 0.231$ \cite{stw}.  We assume a 
spin-zero target.  Finally, the ground-state elastic form factor 
$F(Q^2)$ at momentum transfer $Q$,
\begin{equation}
Q^2=2k^2(1-{\rm cos}\theta),
\end{equation}
is
\begin{equation}
F(Q^2)={1\over Q_w}\int d^3r {{\rm sin}(Qr)\over Qr} [\rho_n(r) - (1 
-4{\rm sin}^2\Theta_W)\rho_p(r)].
\end{equation}
Here $\rho_n(r)$ is the neutron density and $\rho_p(r)$ is the proton 
density.  The form factor is normalized $F(Q^2=0)=1$.
We neglect a small correction from the single-nucleon form 
factors.

The inclusion of $F(Q^2)$ is crucial for heavier targets.  However, 
we evaluate it at relatively small $Q^2$ so the exact form of the 
densities is not important.  The proton density is often well 
constrained by measured charge densities. For simplicity we use theoretical densities from 
simple relativistic-mean-field calculations using the successful 
NL3 effective interaction \cite{NL3}.  These calculations assume spherical 
ground states and do not include pairing corrections.  The use of 
other densities is not expected to change our results significantly.

We now consider a simple ``standard model" for the supernova-neutrino 
spectra, see for example \cite{SNspec}.  This model is close to what 
others have used.  A total energy 
of $3\times 10^{53}$ ergs (1 erg $= 10^{-7}$ J) is assumed to be radiated in neutrinos from a 
supernova at a distance $d$ of 10 kpc ($3.1\times 10^{20}$ m).  For 
simplicity we use Boltzmann spectra at temperatures of $k_BT=3.5$, 5 and 
8 MeV for the $\nu_e$, $\bar\nu_e$ and $\nu_x$ components 
respectively.  Here $k_B$ is the Boltzmann constant, which we set to one in the rest of the paper.  The use of Fermi Dirac spectra (at zero chemical potential) should give similar results.  However, neutrino-nucleus elastic scattering is sensitive to the high-energy tails in the 
spectra.  Therefore non thermal spectra could modify our results 
somewhat and should be investigated in future work.

We assume equal partition in energy among the $\nu_e$ $\bar\nu_e$ 
and the four $\nu_x$ components.  Therefore this standard supernova 
radiates a total of $N_{\nu_e}=3.0\times 10^{57}$, 
$N_{\bar\nu_e}=2.1\times 10^{57}$ and $N_{\nu_x}=5.2\times 10^{57}$ 
neutrinos.  The time integral of the neutrino flux at Earth 
$\phi_i(k)$ for a neutrino of energy $k$ is
\begin{equation}
\phi_i(k)={1\over 4\pi d^2} {N_i\over 2T_i^3} k^2 {\rm e}^{-k/T_i},
\end{equation}
for $i=\nu_e, \bar\nu_e$, or $\nu_x$.  

Microscopic simulations of supernovae suggest that equal partition of energy may be good only to $\approx$ 25 \%.  Furthermore, there is important uncertainty in the $\nu_x$ spectrum, with estimates of $T_{\nu_x}$ ranging from $\approx 6$ to 8 MeV.  It is an important goal of elastic-scattering detectors to measure $T_{\nu_x}$.  Therefore, the predictions of our supernova spectrum have significant uncertainties.  Nevertheless, this simple model should provide order-of-magnitude estimates and may allow easy comparisons to other calculations.

The yield of recoiling nuclei with energy $E$ and mass $M$ is
\begin{equation}
Y(E)= 2\pi N_t \Sigma_{i=\nu_e,\bar\nu_e,\nu_x}\int_0^\infty dk 
\phi_i(k) \int_{-1}^1 d{\rm cos}\theta \delta(E-{Q^2\over 2M}) {d\sigma\over 
d\Omega},
\end{equation} 
where $N_t$ is the total number of target atoms.  For our Boltzmann 
spectra this integral is simple:
\begin{equation}
Y(E)={G^2\over \pi} {Q_w^2\over 4} M F^2(2ME) \bigl({N_t\over 4\pi 
d^2}\bigr) \Sigma_{i=\nu_e,\bar\nu_e,\nu_x} N_i(t_i+1) {\rm e}^{-t_i},
\label{yield}
\end{equation}
with $t_i=(ME/(2T_i^2))^{1/2}$.  For large recoil energy $E$, $Y(E)$ is 
proportional to
\begin{equation}
Y(E) \rightarrow F^2(2ME) {\rm e}^{-({M\over 2 T_{\nu_x}^2})^{1/2} 
E^{1/2}}.
\end{equation}
For light nuclei the high-energy tail continues to hundreds of keV and is 
produced by the scattering of very-high-energy $\nu_x$.  However, for 
heavier nuclei the tail is sharply reduced by the nuclear form factor.

We consider first noble-liquid detectors from $^4$He to $^{132}$Xe 
and then a range of other detectors in order of 
increasing mass number $A$ from $^{12}$C to $^{208}$Pb.  The yield $Y(E)$ from 
Eq. (\ref{yield}) is shown in Figure 1 for detectors made of $^4$He, 
$^{20}$Ne, $^{40}$Ar, $^{84}$Kr, and $^{132}$Xe.  We don't mean 
to imply that detectors would be feasible with all of these liquids.  
However, we show these nuclei to illustrate how the yield depends on 
$A$ for a broad range of $A$.  The spectra in Figure 1 
are peaked at low recoil energy $E$.  Increasing $A$ raises the cross 
section because coherent scattering is proportional to $N^2$.  Thus 
at low energies $Y(E)$ increases significantly with $A$.  However, as 
$A$ increases the spectrum is strongly shifted to lower energies by 
the form factor and the large target mass.  
The energy integral of $Y(E)$, or total yield, is given in Table I in 
events per ton of detector.  Also listed are events above a threshold 
of 5, 10, 25, or 50 keV.  Finally, the average recoil energy of the 
nuclei is given.  This average is influenced by a small number of 
events at high energies, while the spectrum is peaked at low energies.

The optimal choice of $A$ may involve a trade-off between cross 
section, which favors high $A$, and recoil energy, which favors small 
$A$.  This choice may depend on the attainable threshold.  
Furthermore, the choice of target material depends on a host of other 
practical considerations including the presence of possible 
backgrounds from radioactive isotopes.  Although $^4$He has a 
relatively small cross section it has high recoil energies.  We find 
a total yield in Table I of 0.85 events per ton. Helium-based solar-neutrino detectors include HERON \cite{heron} and TPC \cite{tpc} (or HELLAZ \cite{hellaz}).

Perhaps nuclei near $A=40$ give a reasonable balance between cross 
section and recoil energy.  However, Ar may have backgrounds from 
radioactive $^{39}$Ar and $^{42}$Ar, while Kr may have backgrounds 
from $^{85}$Kr.  Xenon is being used in several dark-matter, double-beta decay and solar-neutrino detectors including XMASS \cite{xmass}, XENON \cite{xenon} and ZEPLIN\cite{zeplin}.  The total yield is very large, 31 events per ton.  However, because $^{132}$Xe is heavy, there is a strong premium on obtaining a low threshold.  Background from the $2\nu$ double-beta decay of $^{136}$Xe should not be a problem for a supernova detector.

Above a threshold of 25 keV the yield is a relatively slow function 
of $A$.  Therefore one may have considerable freedom in the choice of 
target material.  We consider the CLEAN liquid-Ne solar-neutrino 
detector \cite{Ne} at some length.  The total yield of 3.99 events 
per ton is dominated by 3.08 $\nu_x$ events with only 0.38 $\nu_e$ 
and 0.53 $\bar\nu_e$ events.  One is very sensitive to the $\nu_x$ 
spectrum.  For example, if the $\nu_x$ temperature is not the 
expected 8 MeV but instead is near that for $\bar\nu_e$ $T_x=5$ MeV, 
the spectrum in Figure 1 is greatly changed.  This verifies that the 
recoil spectrum contains direct information on the $\nu_x$ energies.  
A Monte Carlo simulation of backgrounds in CLEAN is presented in 
section III.

We now consider a number of other targets.  Detectors based on organic scintillator 
such as Borexino and KamLAND \cite{borexino,Kamland} have a yield of 2.50 
events per ton of $^{12}$C, see Table I.  With any carbon-based 
detector there will be backgrounds from $^{14}$C.  Borexino should 
have a ratio $^{14}$C/ $^{12}$C of the order $10^{-18}$ 
\cite{borexino}.  At a concentration of $10^{-18}$ there will be 
about 2 $^{14}$C decays per ton during the 10 seconds of a supernova 
neutrino burst.  This is comparable to the number of elastic 
recoils.  However, the $^{14}$C background should have a different 
spectrum and can be well measured at other times.  Therefore, this 
background may not prevent the use of carbon as a supernova detector, 
even if it does prevent the observation of pp solar neutrinos. 
Quenching may be a serious problem for organic scintillator.  The 
amount of light produced from recoiling C ions may be much less than 
that for recoiling electrons \cite{beacom}.      

An organic scintillator will also have events from neutrino-proton 
elastic scattering \cite{beacom}.  We estimate a yield of about 0.33 
events per ton of CH$_{2}$.  Because of the light proton mass these events will have a larger recoil energy.  The proton elastic-scattering cross section has a theoretical uncertainty of 10 to 20 \% from possible strange quark contributions to the nucleon's axial 
current and spin\cite{strange}.  It would be very useful to have 
better laboratory measurements of neutrino-proton elastic 
scattering.  In contrast, strange quarks are not expected to make 
significant contributions for supernova neutrino-nucleus elastic 
scattering.  Indeed, there is almost no theoretical uncertainty in 
the neutrino-nucleus elastic cross sections.

Silicon detectors such as those described in \cite{cabrera} have a yield of 5.5 events 
per ton and the recoil spectrum is shown in Figure 2, while $^{76}$Ge 
detectors have a yield of 18.6 events per ton and an average recoil 
energy of 9.5 keV.  The double-beta decay experiment Majorana 
\cite{majorana} is proposed to have a 500 kg mass, while Genius 
\cite{genius} has a proposed mass of one ton.  These detectors should 
have very low backgrounds and low thresholds.  Therefore they should 
be sensitive to a supernova at 10 kpc.  It is remarkable that such 
small target masses can yield statistics for our own galaxy 
comparable to those for the historic IMB and Kamiokande signals from SN1987A in 
the Large Magellanic Cloud.

Finally, $^{114}$Cd, $^{130}$Te and $^{208}$Pb yields are listed in 
Table I.  The yield and spectrum for $^{130}$Te is very close to 
$^{132}$Xe (see Figures 1 and 2) since they both have 78 neutrons.  
The heavy nucleus $^{208}$Pb has a very large yield of 47.5 events 
per ton.  However, the average recoil energy is only 2.6 keV.  
Backgrounds and the need for a very low threshold may make a Pb 
detector very difficult to build.  

The nuclei in Figures 1 and 2 display a range of recoil spectra.  For 
point nuclei there would be a single universal spectral shape, with 
the recoil energy decreasing and the yield increasing with 
increasing $A$.  However, the different nuclear form factors modify 
the spectra for heavy nuclei.

Finally, we provide a simple formula to extrapolate the yields in 
Table I to nearby isotopes.  If one ignores small changes in the form 
factors of nearby nuclei, then the yield will be approximately  
proportional to the square of an effective weak charge, $Q_{eff}$,
\begin{equation}
Q_{eff}^2=\delta_{A,odd}3g_a^2 + Q_W^2,
\label{qeff}
\end{equation}
with $Q_W$ from Eq. (\ref{qw}), and $\delta_{A,odd}=1$ for odd $A$ 
nuclei and $\delta_{A,odd}=0$ for even $A$ nuclei.  This factor takes 
into account the axial current of the last nucleon with $g_a=1.26$.  
Because this term adds in quadrature with $Q_W^2$, it makes a very 
small contribution, except for very light systems such as the 
proton\footnote{We ignore the slightly different angular 
distribution of this term.}.  For example, using Eq. (\ref{qeff}) we 
find that $^{21}$Ne and $^{22}$Ne have cross sections respectively 1.29 and 1.48 
times that of $^{20}$Ne.  Natural Ne is 0.3 \% 
$^{21}$Ne and 8.8 \% $^{22}$Ne so this will lead to a slight increase 
in yield over that for pure $^{20}$Ne.

\section{The CLEAN detector}

CLEAN, a detector concept based on liquid Ne, was originally proposed
for the detection of low-energy solar neutrinos.  It will also have
high sensitivity to weakly interacting massive particles (WIMPs). 
Liquid neon has a high scintillation yield, has no long-lived
radioactive isotopes, and can be easily purified using cold traps.  In
addition, neon is inexpensive, dense, and transparent to its own
scintillation light, making it practical for use in a large
self-shielding apparatus.  Here we consider a CLEAN detector in which
a stainless steel tank holds 200 tons of liquid
neon, half of which is exposed to a wavelength shifter to convert the
ultraviolet light to the visible.  Inside the tank and suspended in
the liquid neon are several thousand photomultipliers.  A diagram of
the proposed CLEAN design is shown in Figure 3.

CLEAN will also be sensitive to supernova neutrinos, detected through
neutrino-nuclear scattering.  In this case the entire active neon mass
inside the wavelength shifter can be used, with a possible modest 
fiducial volume cut to reduce radioactive backgrounds.

Of prime importance for determining the sensitivity of CLEAN to
neutrino-nuclear scattering is the determination of the light yield of
liquid neon for nuclear recoils.  Because the density of excitation in
scintillator is typically higher for nuclear recoils than for electron recoils, the chemically
excited species are more likely to interact, increasing the likelihood
that energy will be lost through mechanisms that do not produce light. 
This quality is often expressed as a ``quenching factor", the ratio of
light emitted for a nuclear recoil to the light emitted from an
electron recoil, per unit deposition energy.  While the quenching factor for liquid neon has not yet been measured, we expect it to be similar in magnitude to the quenching factor measured for liquid xenon.  Recent (and widely disagreeing) measurements of the liquid xenon quenching factor\cite{Arneodo,Bernabei} are 22 \% and 43 \%. The amount of quenching for liquid neon should be less than that for liquid xenon, as the density (1.2) of liquid neon is less than that of liquid xenon, while the scintillation mechanism in liquid neon is qualitatively similar.  For the following simulations, we assume that the quenching factor for liquid neon is 0.25. Clearly the scintillation quenching factor in liquid neon would have to be accurately measured in order to properly interpret any supernova data.

As for all neutrino detectors, a prime design consideration in CLEAN
is the reduction of radioactive backgrounds.  We expect that any
radioactive species suspended in the liquid neon will be removed by
passing the neon through charcoal or similar adsorbant; however there
will remain a high rate of gamma rays entering the liquid neon after
being emitted by the surrounding photomultipliers, photomultiplier
support structure, wavelength shifter, and stainless steel tank
containing the liquid neon.

Figure 4 shows the expected supernova neutrino recoil
spectrum for a CLEAN detector with 100 tons of active
liquid neon, assuming 3750 scintillation photons per MeV, 100 \%
efficiency for the wavelength shifter, photomultiplier coverage of 75 \%, and a
photomultiplier quantum efficiency of 15 \%.  Also shown is the expected
radioactive background for 10 seconds of observing time, assuming that
the combined gamma and x-ray emission is dominated by the photomultiplier glass
and the wavelength shifter substrate.  The simulation assumes photomultipliers 20 cm
in diameter, each with mass 650 g, with 30 ng per g of U and Th
and 60 $\mu$g per g of K in the glass.  The wavelength shifter is assumed to be
evaporated on quartz wafers of 1 mm thickness, with a U and Th
concentration of 1 ng per g.  From $10^4$ seconds of simulated data, we find a background of 62 $\pm$ 8 events in 10 seconds within the energy range of 0 to 200 keV.  Here the uncertainty corresponds to the $\pm$ 1-sigma interval for a particular 10-second observation.  Since
liquid neon has no long-lived radioactive isotopes that would limit
its practical threshold, the CLEAN detector could conceivably trigger on as few as two photoelectrons.  The accidental coincidence rate in CLEAN
will be low, as the photomultiplier dark count rate will be suppressed
at liquid-neon temperature (27 K).  Thus we expect that the full 100
tons of liquid neon could be viewed with a threshold of 2
photoelectrons, equivalent to a recoil energy of about 5 keV.  The expected 62 background events is smaller than the expected supernova signal of 330 Ne elastic events above 5 keV in 100 tons.

The radioactive backgrounds, while small in comparison to the
supernova signal, can be lowered further through position resolution,
since most gamma rays will deposit their energy in the outer edges of
the liquid neon sphere.  Recently, Monte Carlo simulations have shown
that the location of ionizing radiation events in CLEAN can be
determined by analyzing the pattern of photomultiplier hits.  These
simulations are described in detail in an upcoming
publication\cite{Coakley}.  Here we show only some results relevant to
supernova neutrino detection.  We find that a mild fiducial radius
cut, leaving a mass of 70 tons, virtually eliminates gamma-ray
background for the purposes of supernova neutrino detection (only 2.6 $\pm$ 1.6
events in the fiducial volume in 10 seconds).  These results are shown
in Figure 5.  In the case of a fiducial volume cut, detecting a few more
photoelectrons will improve the convergence of the position resolution
algorithm.  Currently we can analyze events that produce as few as
8 photoelectrons (recoil energy of 21 keV) if position cuts are
applied.  In general, as the position cut is increased or the number of photoelectrons is reduced, algorithm convergence should be carefully checked.  The expected 2.6 background events are much smaller than the expected supernova signal of 140 Ne events above 21 keV in 70 tons.

The signal-to-noise ratio reported here, though already quite large, might
be significantly improved by the development of photomultipliers with
lower inherent radioactivity.  Such photomultipliers are under
investigation by the CLEAN and XENON collaborations.  In addition,
plans are being made to measure the quenching factor for nuclear
recoils in liquid neon, as this measurement is important for the
determination of the sensitivity of CLEAN to WIMP particles, as well
as for supernova neutrinos.

%In this section we discuss supernova detection in the liquid Neon 
%based CLEAN detector.  After describing the detector, we present 
%expected backgrounds based on a Monte Carlo simulation, discuss 
%possible thresholds and the choice of fiducial volume.  Based on this 
%Monte Carlo simulation, CLEAN looks very promising as a supernova 
%detector.

%Description of detector and its "day job" of low E solar neutrinos.

%Discussion of quenching and the estimate of 0.25.

%Sources of backgrounds.

%Discussion of efficiencies.

%Position reconstruction and fiducial mass.

%Description of Monte Carlo simulation.

%Parameters for the Monte Carlo including a supernova signal duration 
%of 10 seconds.

%Discussion of attainable thresholds, n=8 for position resolution and 
%lower without position resolution...

%Please add a diagram of CLEAN as Figure 3.

%Results for full 100 tons [Figure 4]. What should we say for the 
%position independent threshold?  Between this and 200 keV we predict 
%a total of X background events [Please calculate] and Y[I will 
%calculate] signal events for a supernova at 10 kpc.

%Results for 70 tons [Figure 5].  Between a threshold of 21.5 keV and 
%300 keV we predict a supernova signal of 146 events and a total 
%background of X events [Please calculate].  [Note, my signal number 
%assumes 100 percent efficency.  Should this be changed?]    

%Your summary and conclusions regarding the simulations.

\section{Conclusions}
Detectors with reduced radioactive backgrounds may be
able to study supernovae via neutrino-nucleus elastic scattering. 
This could provide important information on the flux and spectrum of $\nu_x$ ($\nu_\mu$ and $\nu_\tau$).  Elastic-scattering detectors could see a few or more events per ton for a supernova at 10 kpc. This is 20 or more times the number of events per ton of existing water detectors.  The CLEAN experiment, based on the detection of
scintillation in liquid Ne, is a prime example of a detector
that would be sensitive to the elastic scattering of supernova neutrinos.  Its active mass of 100 tons may yield almost 400 $\nu_x$ events.  In addition, many other detectors, including the largest dark-matter and double-beta decay experiments, may also be sensitive to supernova neutrinos via elastic scattering.

Observation of neutrino-nuclear elastic scattering will complement supernova signals from other detectors.  Water detectors such as SNO or Super-K will detect $\nu_x$ without $\nu_x$ spectral information.  This energy information could be important for neutrino oscillations.  KamLAND may be able to measure $\nu-p$ elastic scattering.  This contains energy information just like $\nu-A$ scattering.  The small cross section for $\nu-p$ scattering may be compensated by a large detector mass.  This may yield only slightly smaller statistics than CLEAN.  We strongly encourage development of detectors based on both $\nu-p$ and $\nu-A$ elastic scattering since they may have different backgrounds, thresholds, and systematic errors.  Furthermore, the very large $\nu-A$ elastic cross sections may allow even larger statistics in future detectors.     

\acknowledgments

We thank John Beacom, Steve Elliott, and Harry Miley for useful 
discussions and acknowledge financial support from DOE grant 
DE-FG02-87ER40365 and NSF grant 0226142. We acknowledge the hospitality of the Aspen Center 
for Physics, where some of this work was done.

\vskip 1in

%\begin{tabular}{lllllll} \hline
%\multicolumn{7}{c}{{\bf Table I:} Yield.}   \\
%Target    & Y    & Y $>$ & Y $>$  & Y $>$ & Y $>$ & $<E>$  \\ 
% & & 5 keV & 10 keV & 25 keV & 50 keV & (keV) \\ \hline
%$^4$He &  0.85 & 0.82 & 0.79 & 0.72 & 0.62 & 240  \\
%$^{12}$C & 2.50 & 2.24 & 2.03 & 1.59 & 1.14 & 83 \\
%$^{20}$Ne & 3.99 & 3.32 & 2.85 & 1.95 & 1.17 & 46 \\
%$^{28}$Si & 5.46 & 4.23 & 3.44 & 2.08 & 1.06 & 31 \\
%$^{40}$Ar & 9.39 & 6.56 & 4.95 & 2.49 & 0.99 & 21  \\
%$^{76}$Ge & 18.6 & 9.56 & 5.77 & 1.67 & 0.30 & 9.5 \\
%$^{84}$Kr & 19.8 & 9.49 & 5.46 & 1.40 & 0.20 & 8.4 \\ 
%$^{114}$Cd & 26.3 & 9.71 & 4.60 & 0.70 & 0.041 & 5.7 \\
%$^{130}$Te & 31.8 & 10.14 & 4.30 & 0.47 & 0.014 & 4.8 \\ 
%$^{132}$Xe & 31.1 & 9.78 & 4.09 & 0.43 & 0.012 & 4.8 \\
%$^{208}$Pb & 47.5 & 7.29 & 1.69 & 0.022 & 0.001 & 2.6 \\
%\label{table1}
%\end{tabular}

\begin{tabular}{lllllll} \hline
\multicolumn{7}{c}{{\bf Table I:} Yield.}   \\
Target    & Y    & Y $>$ & Y $>$  & Y $>$ & Y $>$ & $<E>$  \\ 
 & & 5 keV & 10 keV & 25 keV & 50 keV & (keV) \\ \hline
$^4$He &  0.85 & 0.82 & 0.79 & 0.72 & 0.62 & 240  \\
$^{12}$C & 2.5 & 2.2 & 2.0 & 1.6 & 1.1 & 83 \\
$^{20}$Ne & 4.0 & 3.3 & 2.9 & 2.0 & 1.2 & 46 \\
$^{28}$Si & 5.5 & 4.2 & 3.4 & 2.1 & 1.1 & 31 \\
$^{40}$Ar & 9.4 & 6.6 & 5.0 & 2.5 & 0.99 & 21  \\
$^{76}$Ge & 18.6 & 9.6 & 5.8 & 1.7 & 0.30 & 9.5 \\
$^{84}$Kr & 19.8 & 9.5 & 5.5 & 1.4 & 0.20 & 8.4 \\ 
$^{114}$Cd & 26.3 & 9.7 & 4.6 & 0.70 & 0.041 & 5.7 \\
$^{130}$Te & 31.8 & 10.1 & 4.3 & 0.47 & 0.014 & 4.8 \\ 
$^{132}$Xe & 31.1 & 9.8 & 4.1 & 0.43 & 0.012 & 4.8 \\
$^{208}$Pb & 47.5 & 7.3 & 1.7 & 0.022 & 0.001 & 2.6 \\
\label{table1}
\end{tabular}

Yield in events per ton for a supernova at 10 kpc assuming different 
target materials.  Also listed is the number of events above 
thresholds of 5, 10, 25 or 50 keV.  Finally the average recoil energy 
$<E>$ is given.
\vskip 1in

\begin{figure}[h]
\vbox to 4.25in{\vss\hbox to 8in{\hss {\includegraphics{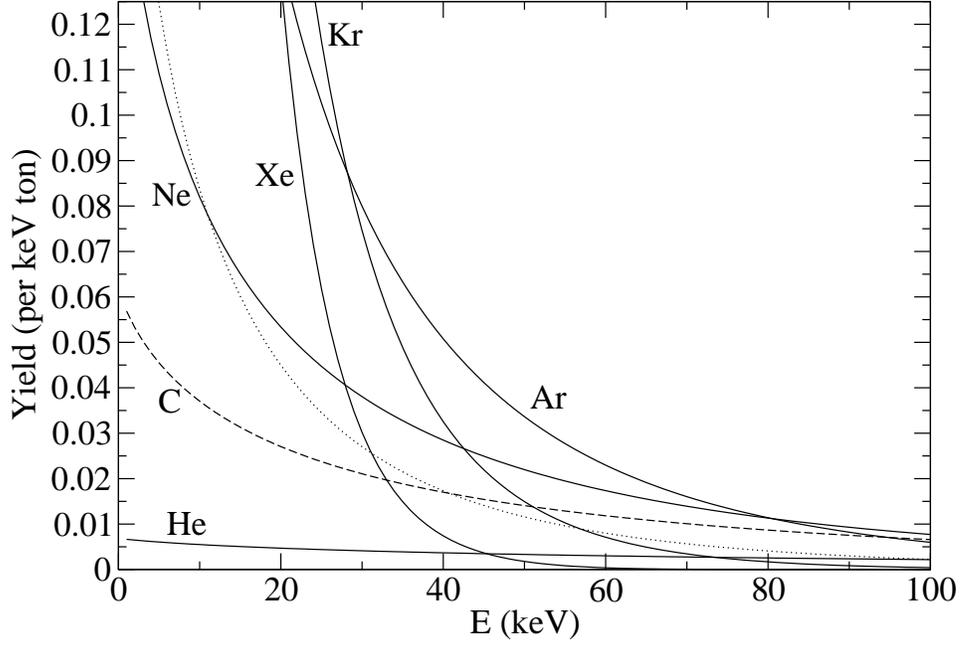}}\hss}
\caption{Yield versus recoil kinetic energy $E$.  The solid curves are 
for noble targets of $^4$He, $^{20}$Ne, $^{40}$Ar, $^{84}$Kr and 
$^{132}$Xe as indicated, the dashed curve is for $^{12}$C.  Finally 
the dotted curve, only shown for $^{20}$Ne, assumes a reduced $\nu_x$ 
temperature of $T_{\nu_x}=5$ MeV.
}
 \label{Fig1}
}
\end{figure}
\begin{figure}[h]
\vbox to 4.25in{\vss\hbox to 8in{\hss{\includegraphics{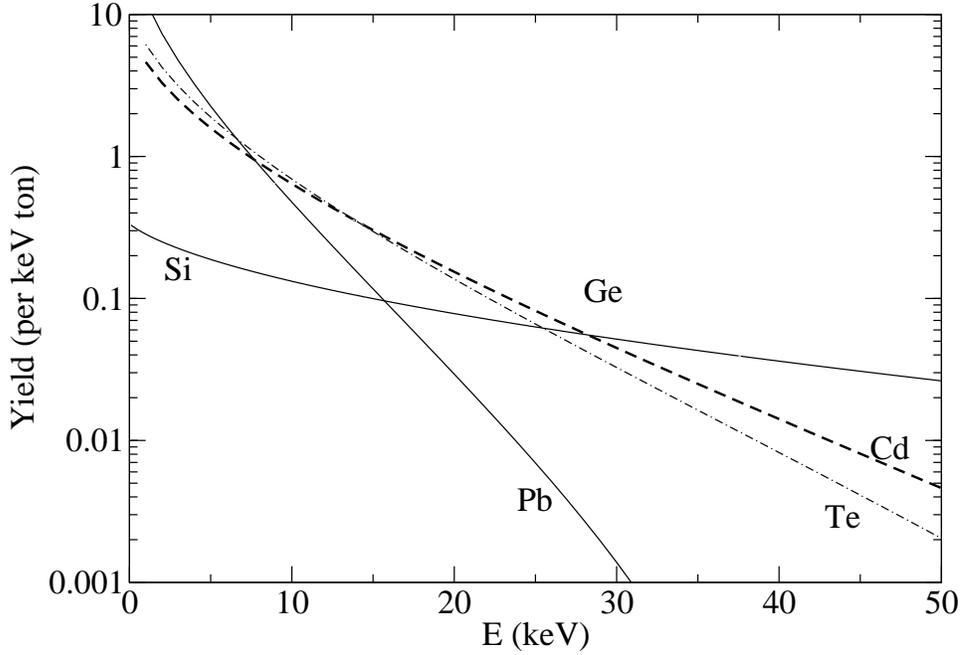}}\hss}
\caption{Yield, versus recoil kinetic energy $E$.  The solid curves are 
for targets of $^{28}$Si and $^{208}$Pb, as indicated, the dashed 
curve is for $^{114}$Cd, the dotted curve $^{76}$Ge, and the 
dot-dashed curve is for $^{130}$Te.  
}
\label{Fig2}
}
\end{figure}

\begin{figure}[h]
\vbox to 4.25in{\vss\hbox to 8in{\hss{\includegraphics{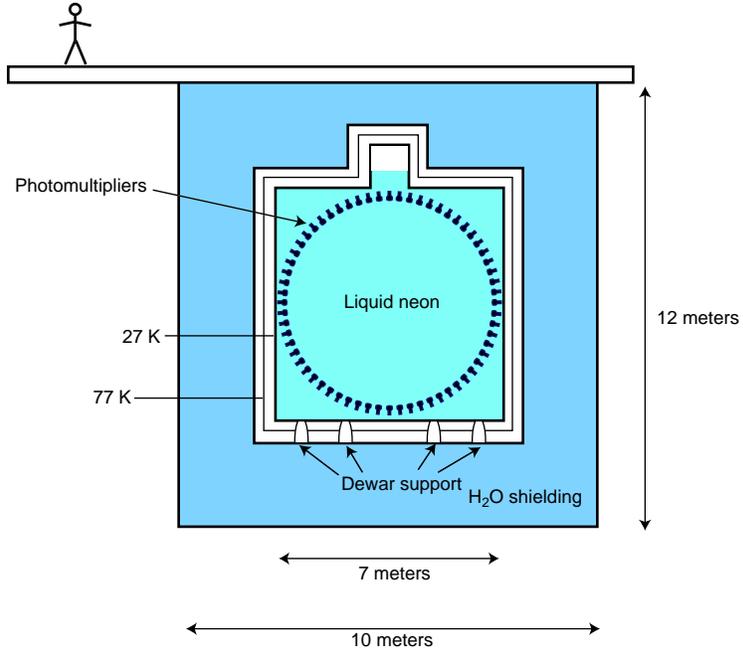}}\hss}
\caption{Diagram of CLEAN}
\label{Fig3}
}
\end{figure}

\begin{figure}[h]
\vbox to 4.25in{\vss\hbox to 8in{\hss{\includegraphics{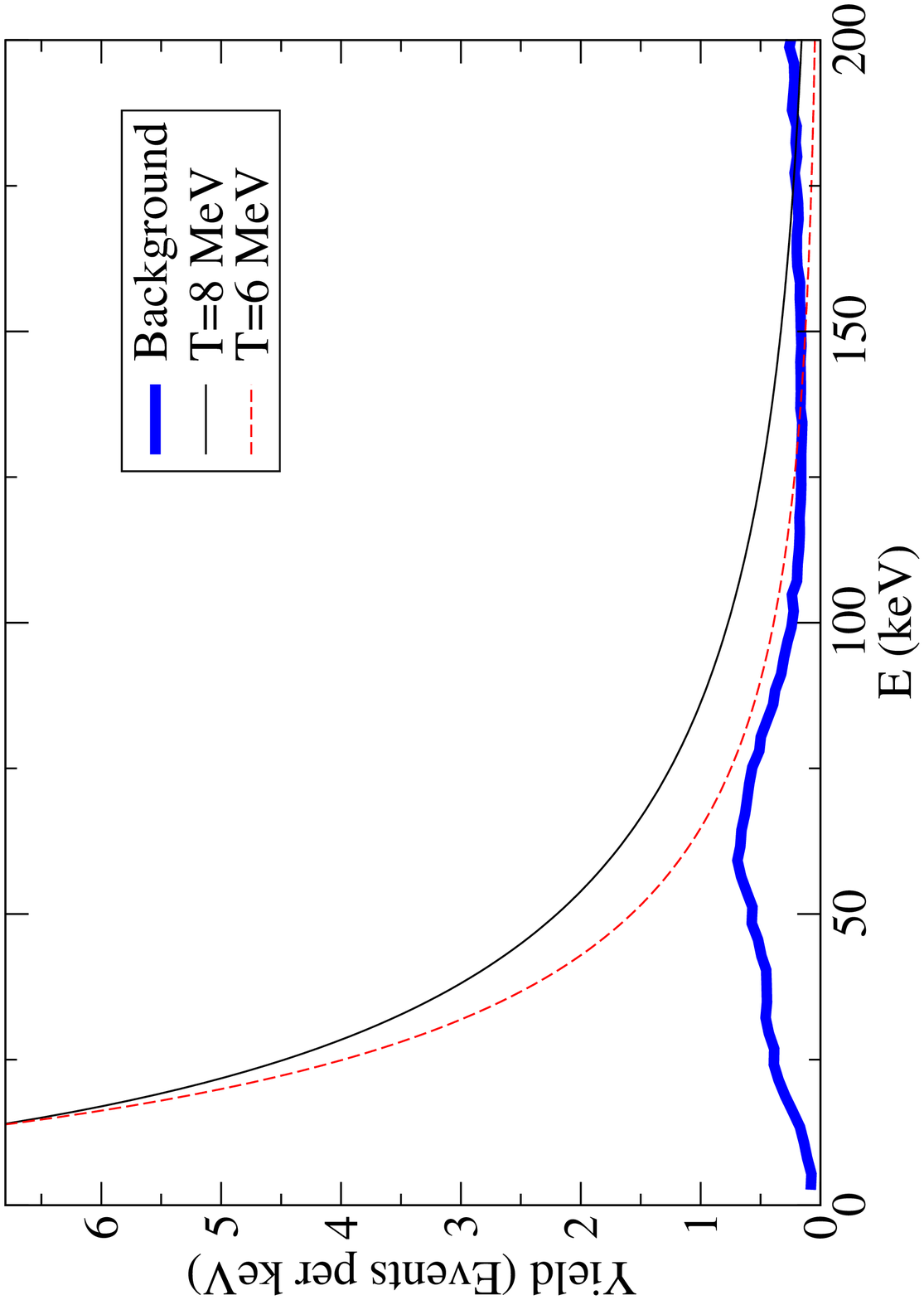}}\hss}
\caption{Yield for full 100 ton fiducial mass of CLEAN versus recoil 
kinetic energy $E$.  The solid curve is the expected supernova signal 
assuming a distance of 10 kpc and a $\nu_x$ temperature $T_{\nu_x}=8$ 
MeV.  The dashed curve assumes $T_{\nu_x}=6$ MeV.  Finally, the thick 
curve is the predicted background from the Monte Carlo simulation assuming an observing time of 10 seconds.  
}
\label{Fig4}
}
\end{figure}

\begin{figure}[h]
\vbox to 4.25in{\vss\hbox to 8in{\hss{\includegraphics{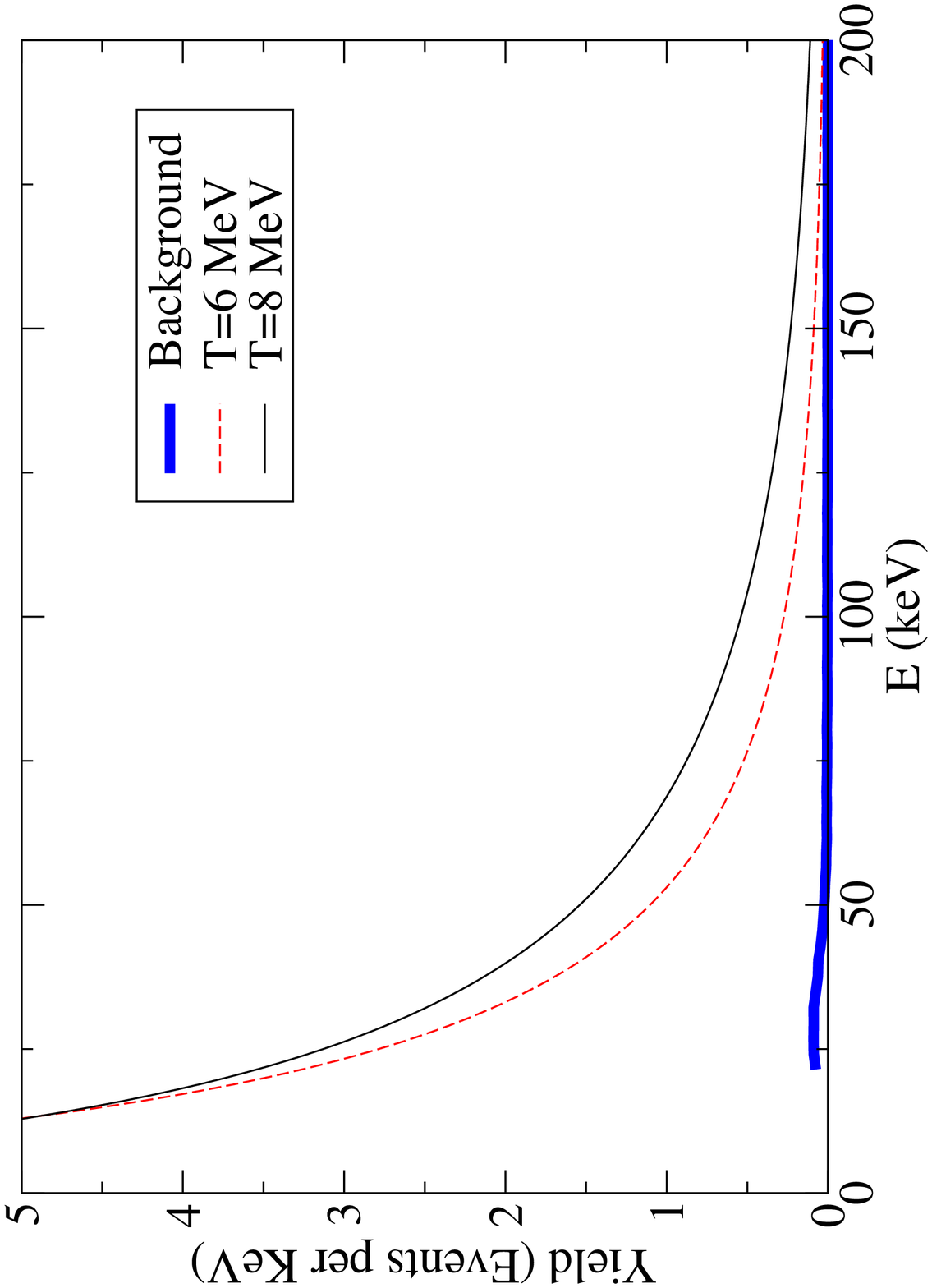}}\hss}
\caption{Yield for 70 ton fiducial mass of CLEAN versus recoil 
kinetic energy $E$.  The solid curve is the expected supernova signal 
assuming a distance of 10 kpc and a $\nu_x$ temperature $T_{\nu_x}=8$ 
MeV.  The dashed curve assumes $T_{\nu_x}=6$ MeV.  Finally, the thick 
curve is the predicted background from the Monte Carlo simulation assuming an observing time of 10 seconds.  
}
\label{Fig5}
}
\end{figure}

\clearpage 


\begin{thebibliography}{}
\bibitem{supernova} See for example, J. F. Beacom, hep-ph/9909231.
\bibitem{SK} Super-kamiokande Collaboration, astro-ph/0007003.
\bibitem{elastic} A. Drukier and L. Stodolsky, Phys. Rev. {\bf D30} 
(1984) 2295.
\bibitem{cabrera} Blas Cabrera, Lawrence M. Krauss, and Frank 
Wilczek, Phys. Rev. Lett. {\bf 55} (1985) 25.
\bibitem{vogel} K. Langanke, P. Vogel and E. Kolbe, Phys. Rev. Lett. {\bf 76} (1996) 2629.
\bibitem{SNO} C. J. Virtue, astro-ph/0103324.
\bibitem{Kamland} V. Barger, D. Marfatia and B. P. Wood, 
hep-ph/0011251.
\bibitem{beacom} John F. Beacom, Will M. Farr, Petr Vogel, Phys.Rev. 
{\bf D66} (2002) 033001.  
\bibitem{elliott} S. R. Elliot, Phys. Rev. {\bf C62} (2000) 065802.  
J. Engel, G. C. McLaughlin, and C. Volpe, hep-ph/0209267.
\bibitem{omnis} D. B. Cline \textit{et al.} Phys. Rev. {\bf D50} (1994) 720; 
P. F. Smith, Astropart. Phys. {\bf 8} (1997) 27; J. J. Zach \textit{et al.}, 
Nucl. Instrum. Meth. {\bf A484} (2002) 194.
\bibitem{land} C. K. Hargrove \textit{et al.}, Astropart. Phys. {\bf 5} (1996) 
183.
\bibitem{raffelt} Mathias Th. Keil, Georg G. Raffelt and Hans-Thomas Janka, astro-ph/0208035.
\bibitem{Ne} D. N. McKinsey and J. M. Doyle, Journal of Low Temperature Physics {\bf 118} (2000) 153.
\bibitem{spectrum} J. D. Lewin and P. F. Smith, Astroparticle Phys. 
{\bf 6} (1996) 87.
\bibitem{CDMS} A. Abusaidi \textit{et al.}, Phys. Rev. Lett. {\bf 84} (2000) 
5699.
\bibitem{hm} L. Baudis \textit{et al.}, Phys. Rev. Lett. {\bf 83} (1999) 41; 
C. E. Aalseth, \textit{et al.}, (IGEX Collaboration), Yad. Phys. {\bf 63} 
(2000) 1299.
\bibitem{majorana} L. DeBraeckeleer, talk at Workshop on the Next Generation U.S. Underground Science Facility, WIPP, June 12-14, 2000, Carlsbad, New Mexico, USA; C. E. Aalseth \textit{et al.}, in Proc. of TAUP'2001, Gran Sasso, Italy, September 2001, ed. A. Bettini \textit{et al.} (2002). See also http://majorana.pnl.gov.
\bibitem{genius}H. V. Klapdor-Kleingrothaus, L. Baudis, G. Heusser, 
B. Majorovits, H. Paes, hep-ph/9910205.
\bibitem{mgd} P. Barbeau, J. I. Collar, J. Miyamoto, and I. Shipsey, hep-ex/0212034.
\bibitem{freedman} D. Z. Freedman, D. L. Tubbs, and D. N. Schramm, 
Annu. Rev. Nucl. Sci. {\bf 27} (1977) 167.
\bibitem{stw} SLD Collaboration, K. Abe \textit{et al.}, Phys. Rev. Lett. {\bf 
86} (2001) 1162.
\bibitem{NL3} G. A. Lalazissis, J. K\"onig and P. Ring, Phys. Rev. 
{\bf C55} (1997) 540.
\bibitem{SNspec} K. Takahashi, M. Watanabe, K. Sato and T. Totani, 
Phys. Rev. {\bf D64} (2001) 093004; T. Totani, K. Sato, H. E. Dalhed 
and J. R. Wilson, Astrophys. J. {\bf 496} (1998) 216.
\bibitem{heron} R. E. Lanou, H. J. Maris, and G. M. Seidel, Phys. 
Rev. Lett. {\bf 58} (1987) 2498; S. R. Bandler et al., J. Low Temp. 
Phys. {\bf 93} (1993) 785.
\bibitem{tpc} G. Bonvicini, D. Naples and V. Paolone, NIM A {\bf 491} (2002) 402; G. Bonvicini and A. Schreiner, NIM A {\bf 493} (2002) 90. 
\bibitem{hellaz} A. de Bellefon for the HELLAZ collaboration, Nucl. 
Phys. B (Proc. Suppl.) {\bf 70} (1999) 386.
\bibitem{xmass} M. Nakahata, http://www.mpi-hd.mpg.de/nubis/www\_lownu2002/transparency/nakahata\_lownu2002.pdf. 
\bibitem{xenon} E. Aprile \textit{et al.}, astro-ph/0207670; Y. Suzuki, astro-ph/0008296.
\bibitem{zeplin} David B. Cline, Hanguo Wang, Y. Seo, astro-ph/0108147.
\bibitem{borexino} BOREXINO Collaboration, G. Alimonti \textit{et al.}, 
Astrop. Phys. {\bf 16} (2002) 205.  Also at hep-ex/0012030.
\bibitem{strange} L. A. Ahrens \textit{et al.}, Phys. Rev. {\bf D35} (1987) 
785.
\bibitem{Arneodo} F.\,Arneodo \textit{et al.}, NIM A {\bf 449} (2000) 
147.
\bibitem{Bernabei} R.\,Bernabei \textit{et al.}, EPJC, to be published.
\bibitem{Coakley}K.\,J.\,Coakley and D. N. McKinsey, to be published.
%\bibitem{weakmag} C. J. Horowitz, Phys. Rev. {\bf D65} (2002) 043001.
\end{thebibliography}
\end{document}